# JOSEPHSON TRAVELING WAVE PARAMETRIC AMPLIFIERS AS NON-CLASSICAL LIGHT SOURCE FOR MICROWAVE QUANTUM ILLUMINATION

*Luca Fasolo* [a,b], *Angelo Greco* [a,b], *Emanuele Enrico* [b], *Fabrizio Illuminati* [c], *Rosario Lo Franco* [d], *David Vitali* [e], *Patrizia Livreri* [d,f,*]

[a] Polytechnic of Torino, Electronic and Telecommunication Department, I-10129 Torino, Italy.
[b] INRiM, Istituto Nazionale di Ricerca Metrologica, I-10135 Torino, Italy.
[c] University of Salerno, Department of Industrial Engineering, I-84084, Fisciano (SA), Italy.
[d] University of Palermo, Department of Engineering, I-90128, Palermo, Italy.
[e] University of Camerino, Physics Division, School of Science and Technology, I-62032, Camerino (MC), Italy.
[f] CNIT, Interuniversity Consortium for Telecommunication, I-56124, Pisa, Italy.
* Corresponding author: patrizia.livreri@unipa.it

***Abstract*** Detection of low-reflectivity objects can be enriched via the so-called quantum illumination procedure. In order that this quantum procedure outperforms classical detection protocols, entangled states of microwave radiation are initially required. In this paper, we discuss the role of Josephson Traveling Wave Parametric Amplifiers (JTWPAs), based on circuit-QED components, as suitable sources of a two-mode squeezed vacuum state, a special signal-idler entangled state. The obtained wide bandwidth makes the JTWPA an ideal candidate for generating quantum radiation in quantum metrology and information processing applications.

## 1. INTRODUCTION

Quantum illumination (QI) is a sensing technique, introduced by S. Lloyd [1] and perfected by S. H. Tan *et al.* [2] in 2008, which exploits quantum entanglement between photons to enhance the detection of low-reflectivity objects immersed in a bright thermal background.

Unlike most of the applications that exploit the quantum properties of matter, such as quantum computing, quantum communication, and quantum cryptography, this detection protocol is superior to its classical counterpart in the presence of noise and decoherence sources. It has been demonstrated [2] that, for a given number $N_S$ of photon composing the detecting signal, in the presence of very low *signal to noise ratios* (SNR ≤ 0.01), this procedure ensures a reduction of 6 dB in the error detection probability ($P_e$) exponent, compared to state-of-the-art classical detection protocol based on coherent light emission from laser or maser, and homodyne detection. In the framework of repeated interrogations with $M$ independent pulses and being $M \gg 1$, it can be demonstrated that with a classical light, it occurs:

$$P_e^{cl} \simeq exp[-MR_{cl}]/2\sqrt{\pi M R_{cl}} \qquad (1)$$

where $R_{cl} = \eta N_S/(4N_B)$ is the classical error rate, η is the reflectance of the transmitted signal on the return path when the target is present, and $N_B$ is the number of background thermal photons with the same frequency of the detecting signal.

Differently, exploiting a quantum light source, the error detection probability can be derived from the so-called *Chernoff bound* [3]:

$$P_e^q \simeq exp[-MR_q]/2\sqrt{\pi M R_q} \qquad (2)$$

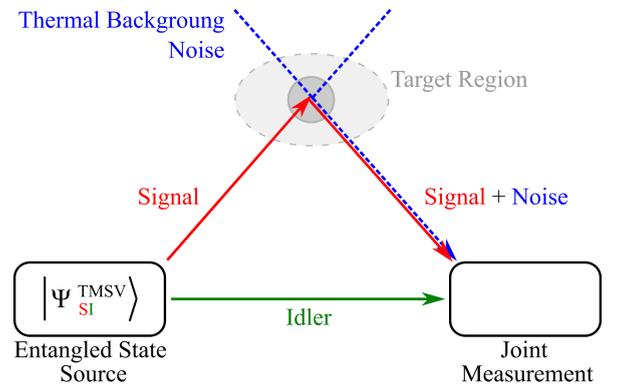

Fig.1: Schematic representation of the quantum illumination protocol

where $R_q = \eta N_S/N_B$ is the quantum error rate and (2) can be experimentally approached exploiting a two-mode squeezed vacuum state (TMSV), which is the optimal in the case of asymmetric discrimination [4] non-classical light sources, and by a proper detection protocol, implementing for example a FeedForward-Sum-Frequency Generation FF-SFG mechanism [5].

From (1) and (2) it can be noted that, for a given $N_S$, an increase of $N_B$ (i.e, a reduction of the SNR) increases the error detection probability in both the protocols, but with a slower rate in the case of the quantum one. This means that the advantage of the quantum protocol with respect to the classical one increases with the reduction of the SNR.

A schematic representation of the quantum illumination protocol is presented in Fig.1. A bi-modal quantum state $|\Psi_{SI}\rangle$ of entangled photons, called respectively *signal* and *idler*, is prepared in the transmitter. The signal mode of this state is sent to the target while the idler mode is temporarily retained at the source. After a certain travel time, the signal mode is partially reflected by the target in the direction of the receiver. Here a joint detection of the reflected signal mode, immersed in the background thermal noise, and of the idler mode is performed. Analysing the correlation between the two photons, now disentangled, one can discriminate between presence or absence of the target.

This general measurement scheme was experimentally tested in the optical range for the first time in 2013 [6] exploiting a beta-barium borate (BBO) non-linear crystal pumped with a laser as source of entangled states and a CCD camera for correlation measurements as a receiver. The measurement scheme was further improved in 2014 [7]. Being the background noise in the optical regime usually negligible, the advantage of the quantum protocol was demonstrated purposely adding an artificial source of noise.

On the contrary, in the microwave regime, the low level SNR condition is naturally realized in a room-temperature environment, making the quantum illumination protocol an essential tool for low-power short-range detection in this spectral region. A first implementation of *Microwave Quantum Illumination* (MQI), also known as *quantum radar*, was proposed for the first time in 2015 [8], exploiting an electro-opto-mechanical (EOM) converter to entangle microwave signal and optical idler fields. The microwave radiation collected from the target was then phase-conjugated and upconverted by another EOM into an optical field for a joint-photodetection at visible wavelengths, a range for which extremely efficient detectors are already available.

It has been proposed and verified [8,9] that the loss of detection efficiency due to the exploitation of the low-quality entangled microwave modes produced by an EOM could be overcome by the use of quantum sources based on Josephson Parametric converter with quantum-limited noise. In 2020 [10], exploiting a Josephson Ring modulator-based source for the MQI and a digital post-processing of the heterodyne revealed signal and idler photons, the detection at room temperature of an object one meter distant from the source was demonstrated (see also [11] where a JPC source was used and compared to a classical noise radar).

A schematic representation of a MQI experimental setup that exploits a Josephson metamaterial as a source of non-classical light is reported in Fig.2. The device is hosted inside a cryostat and feeded with a pump tone generated at room temperature, opportuneously filtered and attenuated. The generated TMSV state is then amplified, both at cryogenic and room temperature, before being splitted. The idler tone is directly sent to the correlation measurement system while the detecting signal is sent in the direction of the target by means of an antenna. The reflected signal, mixed with the background noise, is picked up with a second antenna, amplified and sent to the measurement system.

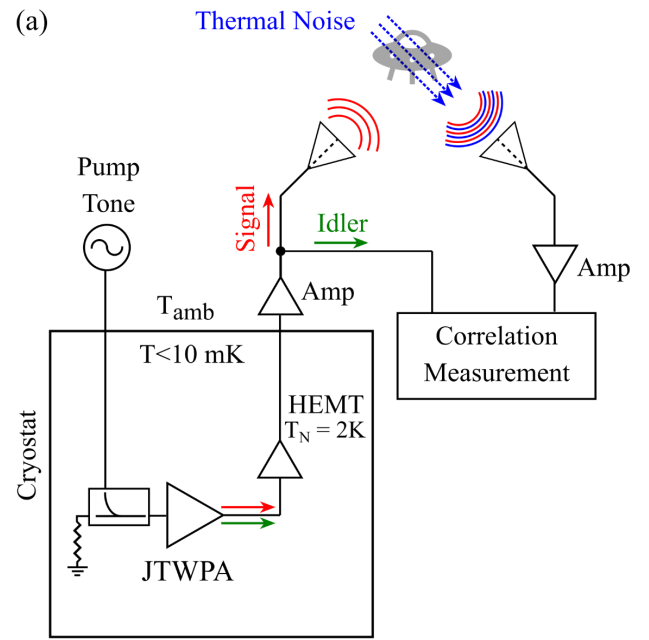

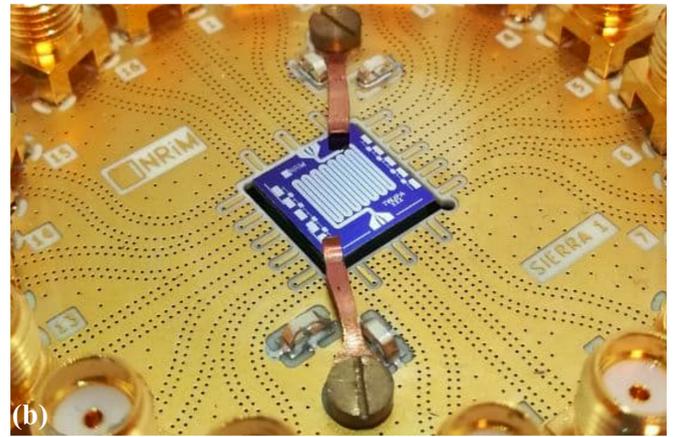

Fig.2: (a) Schematic representation of a MQI protocol that exploits a Josephson metamaterial, hosted in a cryogenic environment, as a source of two-mode squeezed vacuum states for the detection of low-reflectivity targets at room temperature. (b) Micrograph of a typical Josephson Travelling Wave Parametric Amplifier.

The bandwidth of Josephson-based amplifiers or ring modulators is limited by the present technology to tens of megahertz. Much higher bandwidth can be obtained using superconducting metamaterials such as Josephson Traveling Wave Parametric Amplifiers (JTWPA), which achieve bandwidths of several GHz with reasonable gain at the expense of noise figure. [12].

In this context, one can observe that the number of independent pulses $M$ emitted by a light source depends on its bandwidth $W$ and the total integration time $T$ via the simple relation $M = TW$.

It is trivial to appreciate how both classical and quantum detection probabilities depend exponentially by the bandwidth of the light-source. In the following, the bandwidth in which JTWPAs squeeze their output fields and the corresponding capability to generate TMSV states will be discussed.

## 2. JOSEPHSON TRAVELING WAVE PARAMETRIC AMPLIFIER

The Josephson Traveling Wave Parametric Amplifiers (JTWPAs) are on-chip engineered superconducting nonlinear metamaterials for the microwave regime, composed by the repetition of several Josephson junctions (JJs) embedded in a coplanar waveguide. These latter constitute a uniqueness in the solid state physics, being the only known non dissipative nonlinear passive elements. In the context of MQI, these materials can play the same role, as a source of non-classical light, played by the nonlinear crystals in the QI. Indeed, inside these artificial structures the microwave tones can propagate and exchange energy with other tones thanks to intermodulation energy-preserving phenomena, generating entangled quantum states.

The quantum states generated by generic JTWPAs acting as three-wave or four-wave mixers have been extensively investigated [11,13,14,15]. Both working regimes and relative theoretical approaches promote, via parametric amplification, a two-mode squeezed vacuum state $|\Psi_{SI}^{TMSV}\rangle$ thanks to the *non-degenerate spontaneous parametric down-conversion* of a pump photon

$$|\Psi_{SI}^{TMSV}\rangle = \sum_{n=0}^{\infty} \frac{1}{\cosh(\kappa)} (i\tanh(\kappa))^n |n\rangle_S |n\rangle_I \quad (3)$$

Here, the two-mode squeezed vacuum state is expressed in the Fock-state basis of the signal and idler output fields. The parameter $\kappa$ is the modulus of the so-called *squeezing parameter* and it is a quantity that can be expressed as a function both of the characteristics of the components of the metamaterial and of the frequency, thanks to a circuit-Quantum ElectroDynamics (c-QED) description of the system [15].

The squeezing of the two quadratures $q_j$ and $p_j$

$$q_j = a_j + a_j^\dagger, \qquad p_j = i(a_j^\dagger - a_j) \quad (4)$$

of a TMSV state generated by a JTWPA with two different values of $\kappa$ (0.5 and 1.5) are reported in Fig. 3 (respectively (a-b) and (c-d)) in terms of the associated Wigner distributions. Here, $a_j$ and $a_j^\dagger$ represent the bosonic creation and annihilation operators for the $j$-th tone.

It has been experimentally demonstrated how the squeezing parameter $\kappa$ in JTWPA can reach values up to ~3, corresponding to a gain $G$ of about 20 dB [16]. With this in hand, the squeezing capabilities can thus be estimated considering the *squeezing spectrum S* [15,17] in all the bandwidth

Fig. 4 (a) shows the squeezing parameter for three different experimental conditions, which lead to different $\kappa_{MAX}$ in the center of the bandwidth, while Fig. 4 (b) shows the corresponding $S$. From the plot it is clear that the higher the squeezing parameter the more intense the squeezing will be. The squeezing between two quadratures in the low gain case ($\kappa$= 0.5 - 1.5, corresponding to $G$ = 7.2 dB - 1.1 dB) can be appreciated in Fig. 3 and Fig. 4 (b).

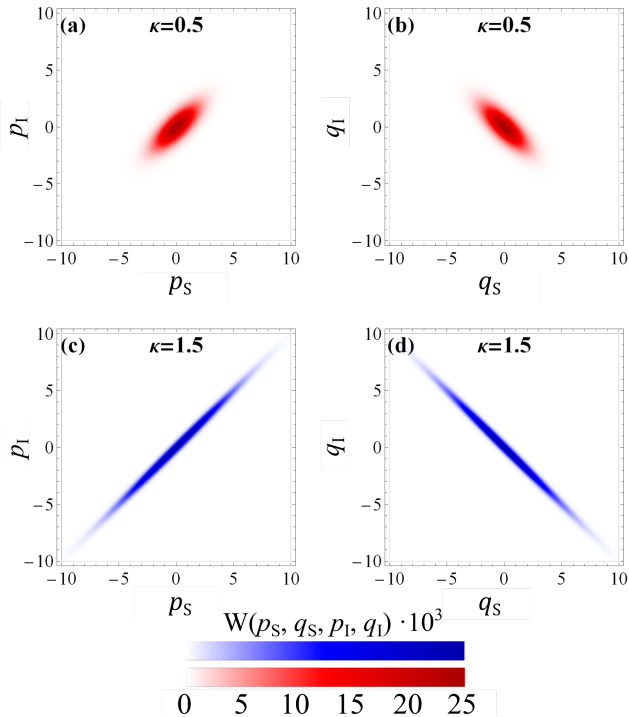

Fig.3: Wigner distributions of the TMSV states for different values of the squeezing parameter. In (a) and (b) $\kappa = 0.5$ while in (c) and (d) $\kappa = 1.5$

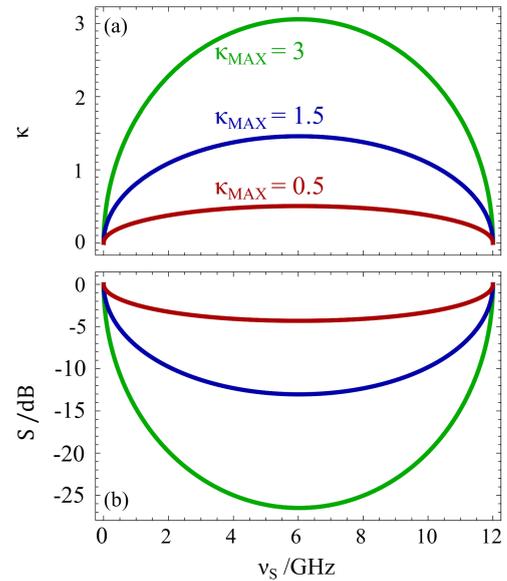

Fig.4:(a) modulus of the squeezing parameter $\kappa$ and (b) squeezing spectrum S as a function of the signal frequency $\nu_S$. The apical values of $\kappa$ for the three curves are 0.5 (red), 1.5 (blue) and 3 (green), leading to a maximum squeezing magnitude of respectively -4.5 dB, -13.2 dB and -26.5 dB. All the curves have been calculated following the model in [15] and supposing a pump frequency of 12 GHz.

## 4. CONCLUSIONS

In this paper, we discussed the role of a JTWPA as an experimental source in the quantum metrology scenario of quantum illumination. In particular, we focused on quantifying its effect on the vacuum state. It emerges that, by suitably controlling the parameters of the device, a broadband two-mode squeezed vacuum state can be generated. This latter can be exploited as a broadband detecting signal that can exponentially increase the performances of MQI protocol, thus making JTWPAs optimal candidate as entangled radiation sources for quantum-enhanced detection of objects in the microwave regime. As a prospect, utilization of these entanglement sources together with manipulation of photon spatial indistinguishability [18], may be strategic for further gains in the detection efficiency against background noise.

## ACKNOWLEDGMENTS

This research has been supported by DARTWARS, a project funded by Istituto Nazionale di Fisica Nucleare (INFN, National Scientific Committee 5), by MUR (Ministero dell'Università e della Ricerca) via the project PRIN 2017 "Taming complexity via QUantum Strategies: a Hybrid Integrated Photonic approach" (QUSHIP) Id. 2017SRNBRK, by the SUPERGALAX project in the framework of the European Union (EU) Horizon 2020 research and innovation programme (H2020, FETOPEN-2018-2020 call), and by the Joint Research Project PARAWAVE of the European Metrology Programme for Innovation and Research (EMPIR). This project (PARAWAVE) received funding from the EMPIR programme co-financed by the Participating States and from the European Union's Horizon 2020 research and innovation programme.